**N₂ accretion, metamorphism of organic nitrogen, or both processes likely contributed to the origin of Pluto's N₂**

Christopher R. Glein

Space Science Division, Space Sector, Southwest Research Institute, 6220 Culebra Road, San Antonio, TX 78238-5166, United States
Email: christopher.glein@swri.org
Phone: 210-522-5510

**Highlights**

- Determining the origin of N₂ is fundamental to understanding the origin and evolution of Pluto.
- The ratio of $^{14}N/^{15}N$ isotopes in N₂ can serve as an important constraint on the source(s) of N₂.
- The currently constrained lower limit on the $^{14}N/^{15}N$ ratio of N₂ on Pluto's surface requires a primordial N₂ or an organic nitrogen source (or both sources) to contribute significantly to the present N₂ inventory of Pluto.
- For these sources to be available, Pluto should have formed from cold planetesimals/pebbles, or a rocky core inside Pluto would have needed to be hot and geologically active.
- A relatively high inferred $^{14}N/^{15}N$ ratio in Pluto's N₂ suggests a diminished role of NH₃ as a N₂ source on Pluto compared with Titan, which points to a key difference in the origin and evolution of these bodies.

**Abstract**

Molecular nitrogen (N₂) plays a profound role in supporting processes on the surface and in the atmosphere of Pluto, yet the origin of Pluto's N₂ remains a mystery. However, this may begin to change as the $^{14}N/^{15}N$ ratio of N₂ was recently estimated based on a non-detection of HC$^{15}$N in Pluto's atmosphere, while accounting for $^{14}N/^{15}N$ fractionation between HCN and N₂ using a photochemical model. Here, I show that, if this latter step of translating isotope ratios is adequately understood, then the derived $^{14}N/^{15}N$ ratio represents the first distinguishing constraint on the origin of Pluto's N₂. One notable finding of the present study is that isotopic fractionation between atmospheric N₂ and N₂-rich ices on the surface of Pluto does not appear to be significant. I infer a lower limit of ~197 for the $^{14}N/^{15}N$ ratio of the dominant (solid) reservoir of N₂ on Pluto; i.e., mostly contained in Sputnik Planitia. From this lower limit, an endmember ammonia source of Pluto's N₂ can be ruled out. I perform N isotope mixing calculations that enable quantitative understanding of the relationships between contributions by primordial N₂, NH₃, and nitrogen originally sourced in organic materials (N$_{org}$) to Pluto's observed N₂ inventory. These calculations also address how uncertainties in the isotopic composition of N$_{org}$ and the history of atmospheric escape affect the allowed ranges of primordial N₂, NH₃, and N$_{org}$ contributions. While present uncertainties are substantial, I find that a contribution by primordial N₂, N$_{org}$, or both is implied, and the sum of their contributions should be at least ~45%. Hence, it is likely that Pluto formed from building blocks that were cold enough to trap N₂ (e.g., <30 K), or Pluto has a thermally processed and dynamic interior that supports generation of N₂ from N$_{org}$ (at temperatures above ~350°C) and N₂ transport to the surface. Furthermore, the lower limit on $^{14}N/^{15}N$ suggests that NH₃ has been a less significant contributor to the origin of N₂ on Pluto than on Titan, which is indicative of a key difference in the origin and evolution of these worlds. Recommendations are given for future work that can continue to advance and contextualize understanding of the origin of Pluto's N₂. A new mission that can determine the origin of $N_2$ on Neptune's moon Triton should be a priority. Such a mission would offer an unprecedented opportunity for comparison of volatile origins on large Kuiper belt objects (past or present) with distinct histories.

**Keywords**

Pluto
Atmospheres, composition
Atmospheres, evolution
Cosmochemistry
Triton

## 1. How did Pluto acquire its molecular nitrogen?

The above is a fundamental question to understand the origin and evolution of Pluto (Singer and Stern, 2015) because $N_2$ is the most abundant gas in Pluto's atmosphere and the dominant observed ice on Pluto's surface (Owen et al., 1993; Gladstone et al., 2016; Grundy et al., 2016a). Moreover, the answer to this question will provide key context for understanding the origin of $N_2$ on other icy worlds in the solar system (e.g., Titan, Triton, Eris; Lunine et al., 1989; Brown, 2012) and possibly beyond the solar system (Xu et al., 2017).

Despite its importance, not much work has been done on this topic as there is little observational data that can be used to constrain the origin of $N_2$ on Pluto. The most detailed study that has been performed since the *New Horizons* flyby of Pluto is the study by Glein and Waite (2018), who suggested that Pluto's $N_2$ might be primordial (i.e., derived from accretion of $N_2$ contained in the building blocks of Pluto). The main support for this scenario is a similarity between the estimated $N_2$ inventory of Pluto and the inventory that would be expected if Pluto accreted icy planetesimals/pebbles with a $N_2/H_2O$ ratio that was similar to that in comet 67P (Glein and Waite, 2018). A potential weakness is that Pluto's surface $CO/N_2$ ratio appears to be much lower than the comet 67P value (Rubin et al., 2019), although this discrepancy could be explained by evolutionary processes that might have plausibly occurred on Pluto (see Neveu et al., 2015; Glein and Waite, 2018). The discovery of $N_2$ in another comet (R2 PanSTARRS) having the highest $N_2$ abundance that has been reported for a comet (McKay et al., 2019; Mousis et al., 2021) expands the range of possibilities for a primordial origin of $N_2$ on Pluto.

There are other options, as reviewed by McKinnon et al. (2021) (see also Scherf et al., 2020). The source of Pluto's $N_2$ could have been $NH_3$, which would have been easier to accrete than $N_2$ (i.e., the condensation temperature of $NH_3$ is higher than that of $N_2$; e.g., Hersant et al., 2004). In addition, the mass balance situation is more than adequate for $NH_3$ (McKinnon et al., 2021), and there is spectroscopic evidence for the availability of ammonia-bearing compounds on Pluto (Dalle Ore et al., 2019). The main issue is whether Pluto's $NH_3$ could have been subjected to energetic processing that is necessary to drive the decomposition of $NH_3$ to $N_2$. The energy source may include ultraviolet light in an early atmosphere, cometary impacts, and hydrothermal activity inside Pluto (see Sekine, 2013; McKinnon et al., 2021).

The other major candidate source material of $N_2$ on Pluto is organic nitrogen ($N_{org}$). McKinnon et al. (2021) showed that Pluto would have started with a large inventory of $N_{org}$ in accreted organic matter if Pluto's building blocks were comet-like (Kissel and Krueger, 1987; Bardyn et al., 2017). It was also shown that the formation of $N_2$ would be thermodynamically favored if organic matter was heated in Pluto's hypothesized rocky core. Abundant methane can also be produced (Kamata et al., 2019; McKinnon et al., 2021), consistent with observations of abundant methane ice deposits on the surface (Moore et al., 2018; Stern et al., 2021). Molecular nitrogen generated from $N_{org}$ or $NH_3$ is not expected to be accompanied by abundant CO, so scenarios involving these source materials do not need to invoke additional processes to explain Pluto's "missing" CO (see McKinnon et al., 2021).

Further study is needed to enable discrimination between primordial and secondary sources of $N_2$ on Pluto. Mandt et al. (2016) and Glein and Waite (2018) proposed that nitrogen isotopes would provide a useful constraint on the origin of Pluto's $N_2$, as already demonstrated at Titan (Mandt et al., 2014; Miller et al., 2019; Erkaev et al., 2020). However, previously the $^{14}N/^{15}N$ ratio of Pluto's $N_2$ was unknown. It has still not been directly measured, but we now have some data that can help move the field forward. Because the N atoms in HCN are derived from $N_2$ via photochemical dissociation, information on the $^{14}$N/$^{15}$N ratio of HCN can be used to constrain the $^{14}$N/$^{15}$N ratio of N$_2$. Indeed, recently Krasnopolsky (2020b) was able to estimate a lower limit on the $^{14}$N/$^{15}$N ratio in atmospheric N$_2$ using a photochemical model and an upper limit on the ratio of $^{15}$N/$^{14}$N in HCN (Lellouch et al., 2017) (see Section 2).

Here, I attempt to take this progress a step further by using Krasnopolsky's (2020b) limiting value as a constraint on the origin of N$_2$ on Pluto. In Section 2, I examine the observationally-based constraint on $^{14}$N/$^{15}$N in N$_2$. Section 3 explores how this constraint can help to clarify the roles of primordial N$_2$, NH$_3$, and N$_{org}$ sources of N$_2$, and their implications for the origin and evolution of Pluto. Section 4 concludes this paper with a summary of conditions that must be satisfied for different N-bearing materials to serve as the source of Pluto's N$_2$, and I give some suggestions for follow-up studies. Note that both $^{15}$N/$^{14}$N and $^{14}$N/$^{15}$N ratios are discussed in this work. The former ratio follows the standard convention in stable isotope geochemistry, while the latter ratio yields numbers that are easier to work with.

## 2. Constraints on the $^{14}$N/$^{15}$N ratio based on observations

Based on the non-detection of HC$^{15}$N in Pluto's atmosphere by the Atacama Large Millimeter/submillimeter Array (ALMA), Lellouch et al. (2017) reported an upper limit on the $^{15}$N/$^{14}$N ratio in HCN of 1/125 (at 2σ). More recent observations give an upper limit of 1/20 (Lellouch et al., 2022). Here, I consider the earlier value that is more stringent. Krasnopolsky (2020b) used his Pluto photochemical model to derive a lower limit of 253 for the $^{14}$N/$^{15}$N ratio of N$_2$ in Pluto's lower atmosphere. This is the only estimate that is available in the published literature, and the value depends on knowledge of isotopic fractionation between N$_2$ and HCN, which is still developing (see below). Krasnopolsky (2020b) also calculated a lower limit for surface N$_2$ ice ($^{14}$N/$^{15}$N > 228), but this value is based on an isotope fractionation factor for NH$_3$. I am not aware of any laboratory measurements of gas-solid isotope fractionation for N$_2$ at Pluto surface temperatures. Nevertheless, Clusius and Schleich (1958) measured the vapor pressures of purified samples of liquid $^{14}$N$_2$ and $^{14}$N$^{15}$N between their normal melting and boiling points (~63-77 K). The vapor pressure isotope effect can be represented by the following equation

$$\log \frac{p^0_{^{14}N_2}}{p^0_{^{14}N^{15}N}} = \frac{0.3985}{T} - 3.43 \times 10^{-3}, \qquad (1)$$

where $p^0_i$ corresponds to the vapor pressure of pure isotopologue $i$, and $T$ designates absolute temperature. Here, I assume that gas-liquid fractionation of N$_2$ can serve as a more accurate proxy of gas-solid fractionation of N$_2$ than does gas-solid fractionation of NH$_3$. I also assume that a modest temperature extrapolation of Equation (1) will not introduce significant inaccuracy (see below).

The gas-solid isotope fractionation factor ($\alpha$) for N$_2$ is defined by

$$\alpha_{N_2, g-s} = \frac{\left(\frac{^{15}N}{^{14}N}\right)_g}{\left(\frac{^{15}N}{^{14}N}\right)_s}. \qquad (2)$$

Assuming that $^{14}$N$^{15}$N mixes ideally with $^{14}$N$_2$ (in terms of Raoult's law) in the condensed phase (see Jancsó, 2004), $\alpha$ can be calculated from the pure component vapor pressures using the following relationship

$$\alpha_{N_2,g-s} = \frac{p^0_{^{14}N^{15}N}}{p^0_{^{14}N_2}}. \tag{3}$$

The estimated value of $\alpha$ at 37 K (i.e., the temperature of equilibrium between atmospheric $N_2$ and $N_2$-rich solid phases on Pluto; Tan and Kargel, 2018) is 0.9832 from Equations (1) and (3). Inserting this value and $(^{14}N/^{15}N)_g > 253$ (Krasnopolsky, 2020b) into Equation (2) yields a lower limit on $^{14}N/^{15}N$ in solid $N_2$ of ~249. It appears that isotopic fractionation between Pluto's atmospheric $N_2$ and $N_2$ ice in equilibrium with the atmosphere is insignificant (~1000 × ln $\alpha_{g-s}$ = −17‰; the solid is enriched in $^{15}N$), compared with the much larger potential effect due to photochemical $N_2$-HCN isotopic fractionation (Krasnopolsky, 2020b).

Alternatively, a semi-empirical approach can be considered to better reflect the uncertainty that comes with translating a $^{14}N/^{15}N$ ratio in HCN to one in $N_2$. Data from Titan can be used as a test since Pluto's photochemistry is analogous to Titan's (Luspay-Kuti et al., 2017; Wong et al., 2017; Krasnopolsky, 2020a). The observed ratios of $^{14}N/^{15}N$ in $N_2$ and HCN on Titan are ~168 (Niemann et al., 2010) and ~72 (Molter et al., 2016), respectively. Krasnopolsky's model for Titan predicts that the $^{14}N/^{15}N$ ratio of HCN should be ~57 (Krasnopolsky, 2016), which is too enriched in $^{15}N$ by a factor of ~1.3. This may suggest that Krasnopolsky's Pluto model also overpredicts the magnitude of isotopic fractionation between $N_2$ and HCN. This effect can be represented by a correction factor ($c$) as shown below

$$\left[\frac{\left(\frac{^{15}N}{^{14}N}\right)_{HCN}}{\left(\frac{^{15}N}{^{14}N}\right)_{N_2}}\right]_{obs} = c \times \left[\frac{\left(\frac{^{15}N}{^{14}N}\right)_{HCN}}{\left(\frac{^{15}N}{^{14}N}\right)_{N_2}}\right]_{model}. \tag{4}$$

I find that on Titan, $c \approx 0.79$ for Krasnopolsky's model. The relatively close consistency between model and observation (obs) can give us some confidence that the model adequately represents photochemical N isotope fractionation. The reason why Krasnopolsky's model slightly overpredicts the enrichment of $^{15}N$ in HCN is because the fractionation factor for $N_2$ predissociation is too large (by up to 38%), or the relative contribution of $N_2$ predissociation to atomic nitrogen production should be smaller than predicted (e.g., 19% rather than 28%). If we assume a similar $c$ value on Pluto, then the $^{14}N/^{15}N$ ratio of Pluto's surface ices would be >197 (or $\delta^{15}N$ less than +380‰, referenced to the standard of Earth's atmospheric $N_2$). While the present approach is simplistic, it has an observational basis and provides a more conservative lower limit than the previously derived value (~249). It is a starting point (see Section 4).

It is reasonable to ask whether Pluto's atmospheric chemistry should be similar to Titan's in producing HCN with a large $^{15}N$ enrichment. A process that typically leads to such enrichment is self-shielding caused by isotopic shifts in the photodissociation cross sections of $N_2$. This results in greater photolytic dissociation of $^{14}N^{15}N$ at lower altitudes (Liang et al., 2007b). The atmosphere needs to be thick enough for this effect to be significant. Pluto's atmosphere may not be thick enough, since it is much thinner than Titan's. Krasnopolsky's (2020b) interpretation based on his analysis of $N_2$ predissociation is that $^{14}N$-$^{15}N$ isotopic fractionation could still be important on Pluto; whereas Mandt et al. (2017) found that photochemistry alone may impart only negligible fractionation, and condensation and aerosol trapping of HCN could be the dominant controls of the $^{14}N/^{15}N$ ratio in HCN on Pluto. Both of these research groups have developed state-of-the-art photochemical models for Pluto consistent with *New Horizons* data, so it is not clear which interpretation should be preferred. I have opted to proceed with Krasnopolsky's (2020b) interpretation because he proposed a specific constraint on the $^{14}$N/$^{15}$N ratio, and it is worth exploring what this would mean for the origin and evolution of N$_2$ on Pluto. Mandt et al.'s (2017) interpretation is also plausible, but quantitative assessment of its implications will have to wait until fractionation factors for HCN condensation and aerosol trapping become available. Because of the disagreement between these models, a source of systematic uncertainty currently exists that should be kept in mind as we proceed.

Below, I will adopt $\left(^{14}\text{N}/^{15}\text{N}\right)_{\text{N}_2,\text{ice}} > 197$ as the current best estimate, and will assume that this constraint applies to the bulk N$_2$ inventory on Pluto's surface. Sputnik Planitia is thought to be the dominant reservoir of surface N$_2$ (Glein and Waite, 2018). Vigorous convective mixing in Sputnik Planitia (McKinnon et al., 2016) can be expected to homogenize its composition of miscible volatiles, making the surface value of the $^{14}$N/$^{15}$N ratio representative of the bulk ice sheet (e.g., McKinnon et al., 2021).

## 3. Comparisons of N isotope ratios

### 3.1. Single source of N$_2$

The inferred lower limit on the $^{14}$N/$^{15}$N ratio of Pluto's surface N$_2$ ice (>197; see Section 2) can be compared to $^{14}$N/$^{15}$N ratios for different origin and evolution scenarios to improve our understanding of the source material(s) of N$_2$ on Pluto. This section seeks to determine if a single source material can be sufficient; in Section 3.2, I consider the possibility of multiple sources.

Now that we have an observational constraint, the next step is to obtain $^{14}$N/$^{15}$N ratios of primordial N$_2$, NH$_3$, and organic nitrogen – the major carriers of N that could have been present in icy planetesimals/pebbles that were accreted by Pluto (see Section 1). Table 1 provides estimated values of the initial $^{14}$N/$^{15}$N ratios for these materials. They are estimates because the $^{14}$N/$^{15}$N ratio has not yet been directly measured in cometary N$_2$, NH$_3$, or bulk organic matter. It is inferred that N$_2$ was the dominant form of N in the solar nebula, with $^{14}$N/$^{15}$N near the solar value (Owen et al., 2001). The $^{14}$N/$^{15}$N ratio of cometary NH$_3$ should be similar to the measured value for NH$_2$ in cometary comae, as NH$_2$ is a dominant photodissociation product of NH$_3$ (Rousselot et al., 2014). Evidence of ammonium salts was recently discovered at comet 67P, and these salts could be the most abundant form of ammonia in comets (Poch et al., 2020; Altwegg et al., 2022). While their isotopic composition is unknown, it is difficult to imagine that their $^{14}$N/$^{15}$N ratios would be much different from that in NH$_3$. Ammonia and ammonium can easily interconvert via proton transfer reactions. For this reason, isotopic distinction between these forms of ammonia is not attempted here. Miller et al. (2019) compiled data on $^{14}$N/$^{15}$N ratios in interplanetary dust particles (IDPs) and *Stardust* organics, but it is not known how many of the IDPs came from comets, or if these small samples are representative of bulk organic matter. Nevertheless, the accreted ratios of $^{14}$N/$^{15}$N listed in Table 1 are thought to be the most reliable values that are currently available, and are consistent with recent literature (Miller et al., 2019).

**Table 1.** $^{14}N/^{15}N$ isotope ratios of major N-bearing materials that could serve as a source of $N_2$ on Pluto.

| Candidate source material of Pluto's $N_2$ | Accreted ratio of $^{14}N/^{15}N$ | $^{14}N/^{15}N$ ratio for enhanced Jeans escape in Pluto's past | Total $^{14}N/^{15}N$ range expected on Pluto if this material is the sole source of Pluto's $N_2$ |
|---|---|---|---|
| Primordial $N_2$ | 441±5 [a] | 327±4 [d] | 323-446 |
| $NH_3$ | 136±6 [b] | 101±4 [d] | 97-142 |
| Organic N | 230±50 [c] | 170±37 [d] | 133-280 |

[a] Marty et al. (2011).
[b] Shinnaka et al. (2016).
[c] Miller et al. (2019).
[d] Calculated using the most likely isotopic enrichment factor (1.35) if the average Jeans escape rate of $N_2$ has been ~$10^4$ times faster than today's rate (Mandt et al., 2016).

Pluto's $N_2$ would have a $^{14}N/^{15}N$ ratio similar to its source if Pluto's $N_2$ experienced minimal atmospheric escape or at least minimal isotopic fractionation during escape. The inferred slow current rate of escape of $N_2$ (Young et al., 2018) may support the former condition. Slow $N_2$ escape rates from Pluto extending into the geologically recent past would be consistent with the lack of a detectable $N_2$ atmosphere on Charon (Stern et al., 2017). Otherwise, too much $N_2$ may be transported to Charon (Tucker et al., 2015). Glein and Waite (2018) showed that the total amount of $N_2$ escape from Pluto would be negligible relative to the observed inventory of $N_2$ if the present escape regime is applicable to the history of Pluto. If past atmospheric escape was significant but occurred through a hydrodynamic mechanism, then there would have been minimal isotopic fractionation. Mandt et al. (2016) predicted that the $^{14}N/^{15}N$ ratio of $N_2$ would only decrease by ~5% for a scenario of hydrodynamic escape.

An alternative is that the structure of Pluto's atmosphere could have been different in the past, with a warmer upper atmosphere supporting a much larger Jeans flux of $N_2$. Such a scenario was also explored by Mandt et al. (2016), who found that the $^{14}N/^{15}N$ ratio of $N_2$ would decrease by a most likely factor of 1.35 relative to the initial $^{14}N/^{15}N$ ratio. Table 1 provides model values of the $^{14}N/^{15}N$ ratio in contemporary $N_2$ if Pluto's $N_2$ was derived from primordial $N_2$, $NH_3$, or $N_{org}$, and was subsequently subjected to a case of enhanced Jeans escape for ~4.5 Gyr. Geological evidence for more extensive past glaciation seems to support this possibility (Moore and McKinnon, 2021). The former glacial ice may have been $N_2$ that was lost to space. We should be mindful, however, that even if past escape rates of $N_2$ were much faster than at present, we do not know whether $N_2$ would have been lost by Jeans escape or hydrodynamic escape, which have very different isotopic consequences (see above).

Figure 1 enables us to deduce which source materials and escape scenarios are consistent with the present lower limit on the ratio of $^{14}N/^{15}N$ in $N_2$ ice on Pluto. It is evident that a $NH_3$ source of Pluto's $N_2$ by itself (see Section 3.2) is inconsistent with the observationally-derived constraint. It does not matter whether the $^{14}N/^{15}N$ ratio has been fractionated by atmospheric escape or not, although the discrepancy is larger when Jeans escape plays a bigger role. I also find that primordial $N_2$ or $N_{org}$ can be solely responsible as the source material for the origin of $N_2$ on Pluto (Figure 1). Primordial $N_2$ always has a greater amount of margin above the lower limit, but considerable uncertainty in the $^{14}N/^{15}N$ ratio of $N_{org}$ (see Section 3.2) permits even its escape-fractionated case (Table 1) to cross above the lower limit.

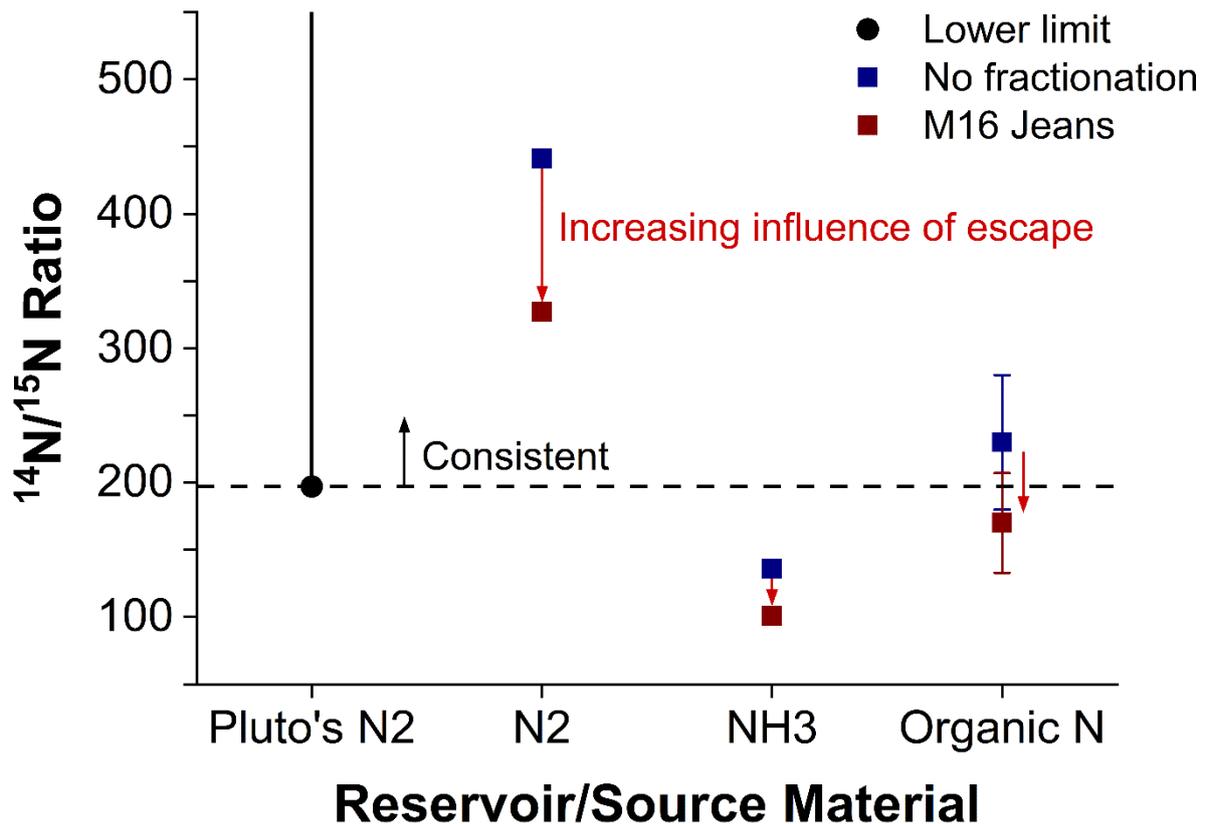

**Figure 1.** Expected contemporary nitrogen isotopic compositions of candidate source materials of Pluto's N$_2$ compared to the present lower limit on the $^{14}$N/$^{15}$N ratio of Pluto's N$_2$ (see Section 2). The squares for each source material represent the accreted ratio (top square) or Mandt et al.'s (2016) most likely $^{15}$N enrichment case for past enhanced Jeans escape (bottom square). Less atmospheric escape than modeled by Mandt et al. (2016) would have an intermediate effect on the $^{14}$N/$^{15}$N ratio in regions between the squares. For a source material to be consistent with the observational constraint, the source material's $^{14}$N/$^{15}$N ratio must lie above the dashed line. Note that uncertainty ranges are not shown for N$_2$ and NH$_3$ since the symbols are larger than the formal error bars. See Table 1 for numerical data.

    Implicit in the preceding analysis is the assumption that chemical reactions producing N$_2$ from NH$_3$ or N$_{org}$ would not exhibit a sizeable isotope effect. For endogenic production, this is very likely to be true. Li et al. (2009) experimentally showed that thermal decomposition of NH$_3$ yields a kinetic fractionation factor of 0.983. This degree of isotopic fractionation (~2%) is minuscule compared with the differences between cases in Table 1. I am not aware of a mechanistic description of how N$_2$ is formed from N$_{org}$, but since two nitrogen atoms need to come together, it seems reasonable to assume that N would first be released from organic matter into a fluid phase. Experiments demonstrate that free ammonia can be released when organic matter from carbonaceous chondrites is subjected to hydrothermal conditions (Pizzarello and Williams, 2012). Once NH$_3$ is released, it may undergo thermal decomposition to N$_2$, with a fractionation factor similar to the previously mentioned one.

    A different possibility is that NH$_3$ photolysis in an early atmosphere on Pluto could have generated N$_2$ (e.g., Atreya et al., 1978). In this case, the isotopic consequences are more interesting, although they do not change my conclusion about the general inconsistency of a NH$_3$ source of N$_2$ and may strengthen this conclusion instead. Liang et al. (2007a) used measured photoabsorption

cross sections of $^{15}NH_3$ and $^{14}NH_3$ to model isotopic fractionation associated with ammonia photolysis in Jupiter's upper atmosphere. They found that $^{15}NH_3$ is destroyed faster than $^{14}NH_3$. This means that photochemically-generated $N_2$ is expected to have a lower $^{14}N/^{15}N$ ratio than the starting $NH_3$. If this finding is applicable to Pluto, then a photochemical $NH_3$ source of $N_2$ would fall on or below the $NH_3$ points shown in Figure 1, perhaps down by a factor of ~0.9 as suggested by the results of Liang et al. (2007a).

For simplicity's sake, the following discussion assumes that Pluto's $N_2$ had a single source material. If this is true, then $NH_3$ cannot be the source; otherwise, the $^{14}N/^{15}N$ ratio of Pluto's $N_2$ would be too low. An exception could be if $N_2$-rich ices with low $^{14}N/^{15}N$ ratios are buried in Sputnik Planitia. However, such compositional layering with less volatile species at greater depths has been argued to be unlikely (see McKinnon et al., 2021). The present interpretation imposes constraints on past environments on Pluto. For example, to prevent efficient photochemical formation of $N_2$ from $NH_3$, Pluto's surface temperature should not have exceeded ~150 K (Atreya et al., 1978). The rate of accretional heating should have been relatively weak to keep surface temperatures from getting too high. Thus, Pluto probably did not rapidly accrete lots of large (km-scale) planetesimals. Pebble accretion seems to be the favored formation process, as it is for many other reasons (see McKinnon et al., 2021). A massive greenhouse atmosphere (e.g., Lunine and Nolan, 1992) should also be avoided. Another potential implication is that, if Pluto has/had hydrothermal systems at the base of a subsurface water ocean (Nimmo et al., 2016; Johnson et al., 2016), they should not be overly hot or oxidized. Cooler, more reduced ($H_2$-rich) systems would inhibit the decomposition of $NH_3$ to $N_2$ (Glein et al., 2008; 2009), favoring the preservation of ammonia, which is at least qualitatively consistent with the presence of ammonia-bearing compounds on the surface of Pluto (Dalle Ore et al., 2019). Such conditions would favor the conversion of $CO_2$ to $CH_4$, which could help to explain the ubiquity of methane ice on Pluto's surface and the mysterious apparent absence of $CO_2$ ice (Grundy et al., 2016a).

The consistency of primordial $N_2$ or organic N to serve as the sole source of Pluto's $N_2$ suggests that Pluto formed from cold materials, or Pluto's deep interior was hot. Accretion of $N_2$ can only occur at low temperatures (e.g., <30 K) because of the high volatility of $N_2$ (Rubin et al., 2015). If Pluto's $N_2$ is primordial, then Pluto's building blocks could not have formed too close to the Sun, even though Pluto is thought to have migrated outward (see McKinnon et al., 2021). An additional implication is that $CH_4$ should be accreted if $N_2$ is accreted, since $CH_4$ has a higher condensation temperature than that of $N_2$ (e.g., Hersant et al., 2004). Pluto's methane could also be a primordial species. This logic applies to CO as well, but in this case, consistency with observational data requires a mechanism to keep CO from being more abundant than $N_2$ on the surface (see Glein and Waite, 2018). In contrast, high temperatures ($\gtrsim$350°C) in a rocky core are required to decompose organic matter and produce $N_2$ from $N_{org}$ (Miller et al., 2019). The presence of $N_2$ on Pluto's surface that was derived from organic metamorphism would imply that Pluto underwent differentiation and has a rocky core (Nimmo and McKinnon, 2021; Denton et al., 2021), and large-scale mass transfer from the core to the surface has occurred (which may be linked to identified cryovolcanic features; Singer et al., 2022). In the latter case, an appreciable amount of radiogenic $^{40}Ar$ may also be transported from the core to the surface, depending on when core volatiles were delivered to the surface. Core volatiles may not have been produced early in Pluto's history (e.g., earlier than ~1 Gyr after Pluto's formation), because it would have taken time for the core to warm up enough so that organic matter can be cracked (Kamata et al., 2019). This heating timescale is intriguingly similar to the half-life of radioactive $^{40}K$ (1.25 Gyr; Steiger and Jäger, 1977). Strictly speaking, a large inventory of surface $^{40}Ar$ on Pluto would represent compelling evidence of volatile transport from the core, while a lack of $^{40}Ar$ would not necessarily mean that core volatiles were not delivered to the surface.

Lastly, it should be noted that N$_2$ generation from N$_{org}$ is not expected to depend on redox conditions, owing to a likely deficiency of hydrogen atoms in the core that limits NH$_3$ production (McKinnon et al., 2021).

*3.2. Mixed source of N$_2$*

Another possibility is that multiple materials contributed to the origin of N$_2$ on Pluto. This is not required by current data, which permits primordial N$_2$ or N$_{org}$ as the only source material of Pluto's N$_2$ (Figure 1). If simplicity is favored (e.g., Owen and Encrenaz, 2006), then the interpretation in Section 3.1 may be most relevant. On the other hand, the existing data are very limited, and if we treat Titan's atmospheric N$_2$ as a guide for Pluto, then a mixed source might be more realistic. Miller et al. (2019) found that the ratios of $^{15}$N/$^{14}$N and $^{36}$Ar/$^{14}$N on Titan fall close to a mixing line between NH$_3$ and N$_{org}$ sources of N$_2$. A significant contribution from primordial N$_2$ is ruled out because the abundance of primordial $^{36}$Ar is very low in Titan's atmosphere (Niemann et al., 2010). On Titan, binary source mixtures of NH$_3$ and N$_{org}$ can reproduce the observational data, with inferred nominal contributions of ~50% of N$_2$ from NH$_3$ and ~50% of N$_2$ from N$_{org}$ (Miller et al., 2019). The goal of this section is to determine what kinds of mixtures would be consistent with the isotopic constraint for Pluto from Section 2.

A mixing equation is needed to calculate the N isotopic composition of N$_2$ derived from primordial N$_2$, NH$_3$, and organic nitrogen. The following equation is used for this purpose in the present paper

$$R_{\text{mix}} = f_{N_2} R_{N_2} + f_{NH_3} R_{NH_3} + f_{N_{org}} R_{N_{org}}, \tag{5}$$

where $R_i$ represents the $^{15}$N/$^{14}$N ratio of source material (or mixture) $i$, and $f_i$ stands for the fraction of nitrogen that is contributed by source material $i$ to the mixture (mix). Note that the isotopic composition is expressed in terms of the $^{15}$N/$^{14}$N ratio rather than the $^{14}$N/$^{15}$N ratio in this equation. When the equation is set up in this way, the fractions technically refer to the fraction of $^{14}$N, which is almost identical to the fraction of total N since $^{15}$N is always much less abundant than $^{14}$N. Here, I evaluate Equation (5) using the two sets of 1/$R$ values in Table 1. These correspond to scenarios in which there has not been appreciable evolution of the $^{14}$N/$^{15}$N ratio of N$_2$ in the surface-atmosphere system on Pluto, or atmospheric escape has led to significant modification of the $^{14}$N/$^{15}$N ratio (see Section 3.1). In the latter scenario, the source materials would contribute a lower $^{14}$N/$^{15}$N ratio to the present-day mixture, as shown in Table 1. One additional equation is coupled to Equation (5) to constrain the fractional contributions from primordial N$_2$, NH$_3$, and N$_{org}$. The mass balance for this system is given by

$$f_{N_2} + f_{NH_3} + f_{N_{org}} = 1. \tag{6}$$

The $^{15}$N/$^{14}$N ratio of various mixtures can now be calculated as a function of any two of these $f$ variables. Simple inversion then yields $^{14}$N/$^{15}$N ratios.

There are different ways to examine mixing relationships of N isotopes on Pluto. Two examples that provide a high-level perspective are shown in Figure 2 (note, a plot of this type was first presented by Miller et al., 2017). It can be seen in Figure 2a that a large number of $f$ value combinations are consistent with the present observational constraint; the system is degenerate. Nevertheless, it is more useful to have a lower limit on the $^{14}$N/$^{15}$N ratio than no data at all. We can exclude some regions of the parameter space, which represents progress. A comparison between Figure 2a and 2b reveals that the region of consistency below the dashed lines shrinks if atmospheric escape has caused an enrichment of $^{14}N^{15}N$ in retained $N_2$. In other words, the parameter space would be more constrained if escape has been important. For both cases, it can be concluded that there needs to be a sizeable (>45%) contribution by primordial $N_2$, organic N, or both materials. Ammonia alone is an insufficient isotopic source of $N_2$ (see Section 3.1). However, when considering a mixed source, we must avoid the temptation to assume that $NH_3$ has played a minimal role in the origin of Pluto's $N_2$. As an example, Figure 2a shows that $NH_3$ could contribute as much as ~55% of the nitrogen atoms in today's $N_2$. Conversely, for the case where escape is important, only a minor (<29%) contribution by $NH_3$ is allowed, but not necessarily zero (Figure 2b). A non-zero (>29%) contribution by primordial $N_2$ appears to be implied in this case to provide a large enough increase in the $^{14}N/^{15}N$ ratio of the mixture.

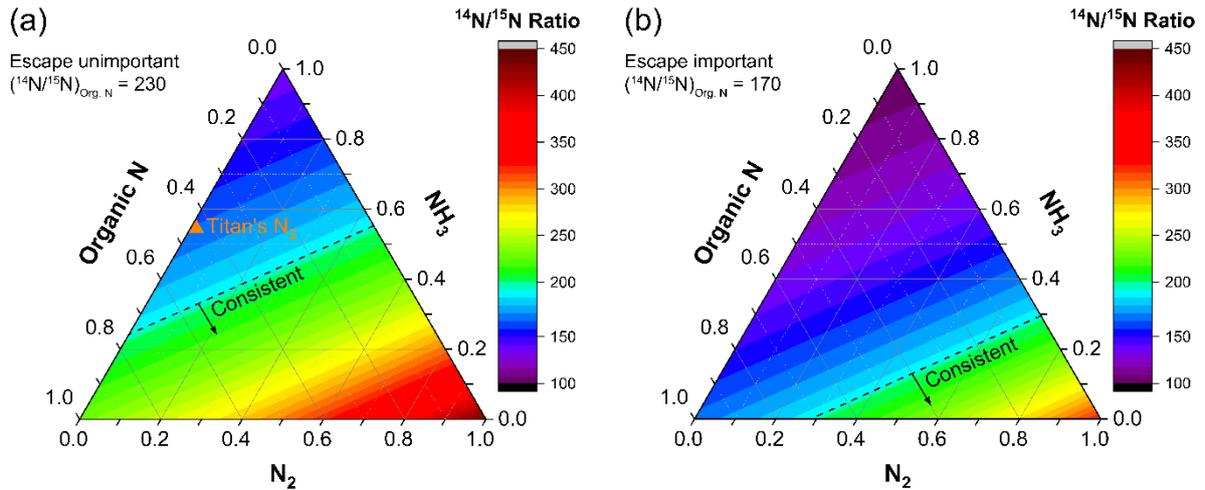

**Figure 2.** Ternary contour plots showing the parameter space for the bulk $^{14}N/^{15}N$ ratio from mixing of primordial $N_2$, $NH_3$, and organic nitrogen. The axes refer to the fractional contributions of the indicated source material. Plot (a) would apply to Pluto's $N_2$ if the $^{14}N/^{15}N$ ratio of Pluto's $N_2$ directly reflects that of its source(s) (e.g., if Pluto's $N_2$ has not experienced significant isotopic fractionation due to atmospheric escape). Plot (b) (with isotope ratios for the components taken from the third column in Table 1) would be more representative if Jeans escape of $N_2$ was greatly enhanced in Pluto's past. The dashed line in both plots indicates the present lower limit on the $^{14}N/^{15}N$ ratio of $N_2$ ice on Pluto (see Section 2). Mixtures that lie below the dashed line can provide acceptable solutions to the origin of $N_2$ problem on Pluto. The orange triangle in plot (a) denotes Miller et al.'s (2019) proposed solution for the origin of $N_2$ on Titan. In both plots, midrange values for the $^{14}N/^{15}N$ ratio from organic N are adopted.

A different view of the mixing relationships is shown in Figure 3. This perspective is more useful for understanding how the main sources of uncertainty affect the allowed ranges of $f_{N_2}$, $f_{NH_3}$, and $f_{N_{org}}$. It can be seen that the amount of $N_2$ derived from $NH_3$ must always be less than 100%, and even lower for cases in which the N isotopic composition has undergone substantial evolution because of atmospheric escape. Uncertainty in the $^{14}N/^{15}N$ ratio of an organic source of $N_2$ can also play a pivotal role. If organic N had a sufficiently high $^{14}N/^{15}N$ ratio, then a contribution by primordial $N_2$ is not needed (Figure 3b, 3c, 3f). Otherwise, at least some primordial $N_2$ is needed. The lower limit on $f_{N_2}$ increases with decreasing $^{14}N/^{15}N$ ratio of accreted organic matter, and can be increased further if atmospheric escape led to a significant decrease in the $^{14}N/^{15}N$ ratio of Pluto's $N_2$. The upper limit on the fraction of organic-derived $N_2$ responds in an opposite manner to these changes. The limiting values obtained from this analysis are given in Table 2. I find that specific scenarios can be constrained to varying degrees, but currently the data only allow a meaningful total upper limit to be placed on the contribution by $NH_3$.

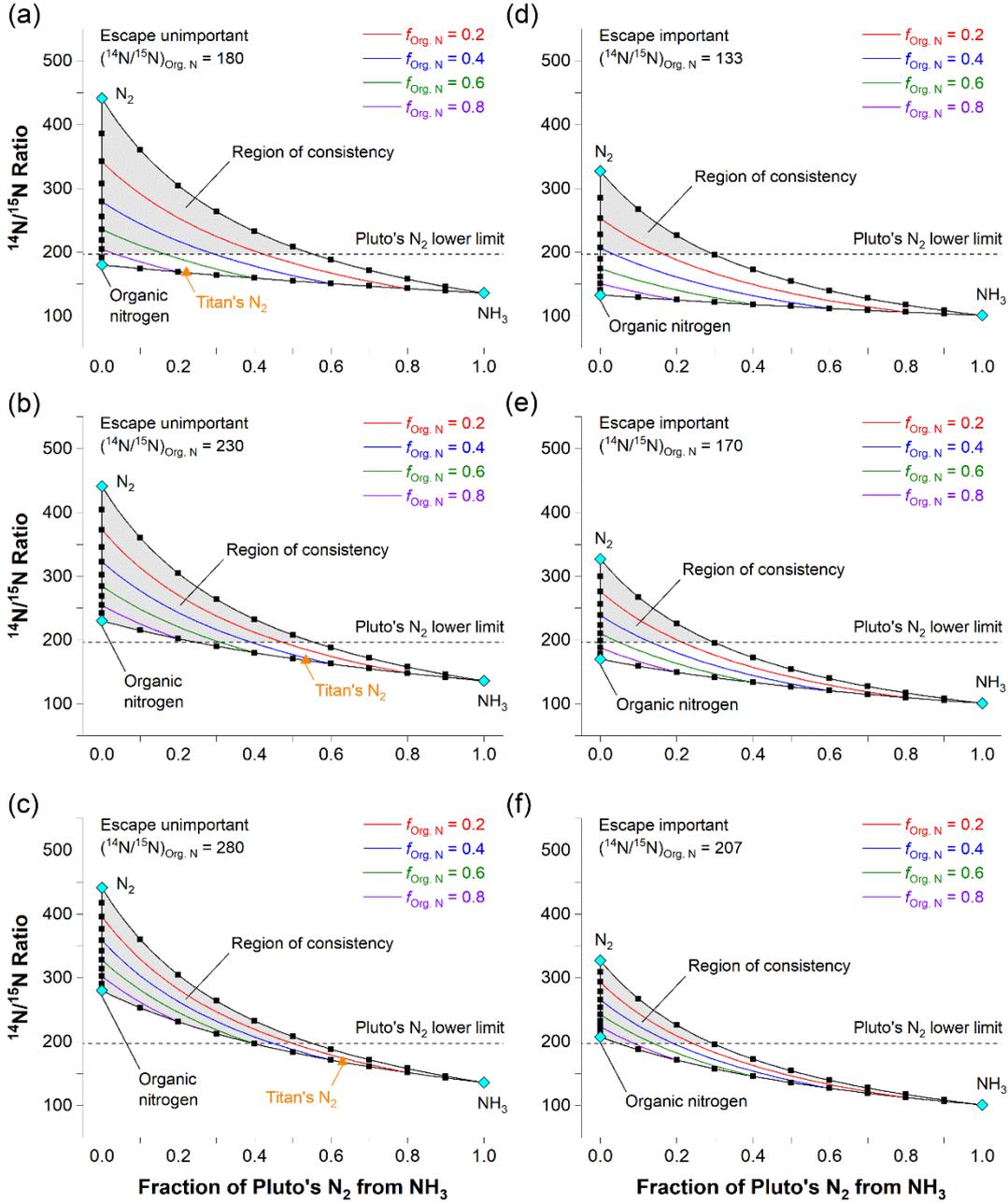

**Figure 3.** $^{14}N/^{15}N$ ratio of mixtures of primordial $N_2$, $NH_3$, and organic nitrogen on Pluto. Vertical plots demonstrate how uncertainty in the $^{14}N/^{15}N$ ratio of organic N (Table 1) affects the $^{14}N/^{15}N$ ratio of the mixture. Side-by-side plots show results for models in which there has been minimal evolution of the N isotopic composition of $N_2$ in the surface-atmosphere system on Pluto (left plots); or significant fractionation of N isotopes has occurred, mainly as a result of enhanced Jeans escape over the history of Pluto (right plots). In each plot, cyan diamonds indicate the endmember components, black curves correspond to binary mixtures, black squares represent 10% changes in the binary composition, and colored curves give $^{14}N/^{15}N$ ratios at fixed values of the fractional contribution (*f*) provided by organic N. Note that the top curve in each plot has no $N_2$ that comes from organic N, and the bottom curve in each plot has no $N_2$ that comes from primordial $N_2$. The dashed line indicates the present lower limit on the $^{14}N/^{15}N$ ratio of $N_2$ ice on Pluto (see Section 2).

Mixtures that lie above the dashed line are consistent with this constraint, and the corresponding parameter space is shaded in gray. The orange triangles in plots (a), (b), and (c) denote mixture compositions that can reproduce observed ratios of $^{15}$N/$^{14}$N and $^{36}$Ar/$^{14}$N in Titan's atmosphere (Miller et al., 2019).

**Table 2.** Summary of the amounts that different N-bearing precursors can contribute to Pluto's present surface inventory of N$_2$. The fractional contributions (*f*) depend on whether atmospheric escape processes have significantly fractionated N isotopes over Pluto's history (see Section 3.1), and they also depend on the N isotope ratio that could be contributed by accreted organic matter (Table 1). For a given combination of these two variables, the range in *f* for each precursor is deduced that would allow the mixed N isotopic composition to be consistent with the estimated lower limit on $^{14}$N/$^{15}$N in N$_2$ ice on the surface of Pluto (see Section 2). The final row gives the most conservative ranges, assuming that we know neither the nature of the integrated escape regime, nor the value of $\left(^{14}\text{N}/^{15}\text{N}\right)_{\text{N}_{\text{org}}}$ with strong certainty.

| Integrated escape regime | $\left(^{14}\text{N}/^{15}\text{N}\right)_{\text{N}_{\text{org}}}$ | Allowed range for $f_{\text{N}_2}$ | Allowed range for $f_{\text{NH}_3}$ | Allowed range for $f_{\text{N}_{\text{org}}}$ |
|---|---|---|---|---|
| Today's Jeans or hydrodynamic [a] | 180 | 0.15-1 | 0-0.55 | 0-0.85 |
| Today's Jeans or hydrodynamic [a] | 230 | 0-1 | 0-0.55 | 0-1 |
| Today's Jeans or hydrodynamic [a] | 280 | 0-1 | 0-0.55 | 0-1 |
| Enhanced Jeans [b] | 133 | 0.55-1 | 0-0.29 | 0-0.45 |
| Enhanced Jeans [b] | 170 | 0.29-1 | 0-0.29 | 0-0.71 |
| Enhanced Jeans [b] | 207 | 0-1 | 0-0.29 | 0-1 |
| | Total ranges: | 0-1 | 0-0.55 | 0-1 |

[a] Negligible evolution of the $^{14}$N/$^{15}$N ratio in N$_2$ is assumed (1) because the quantity of N$_2$ that has escaped from Pluto would have been negligible if today's Jeans regime is extrapolated into the past (Glein and Waite, 2018), or (2) because hydrodynamic escape is only weakly dependent on the masses of N isotopes (Mandt et al., 2016).
[b] N isotope fractionation based on Mandt et al.'s (2016) model of atmospheric evolution with a large Jeans escape rate for N$_2$ of ~1×10$^{27}$ molecules/s.

Atmospheric N$_2$ on Titan provides a relevant point of comparison (Figures 2a; 3a, 3b, 3c). It has a bulk $^{14}$N/$^{15}$N ratio of 167.7±0.6 (Niemann et al., 2010), which is below the lower limit for Pluto's N$_2$ ice of ~197 (see Section 2). This signifies a key difference between the sources of N$_2$ on Titan and Pluto. Note that a point rather than a line is shown for Titan's N$_2$ in these plots, as the measured ratio of $^{36}$Ar/N$_2$ allows a unique solution to be derived for the proportions of source materials on Titan (Miller et al., 2019). Could Pluto's N$_2$ have been formed from the same source materials as those inferred on Titan (NH$_3$ and N$_{\text{org}}$), but with different contributions? Figure 2a shows that such a scenario is possible if $f_{\text{NH}_3} < 0.24$ and $f_{\text{N}_{\text{org}}} > 0.76$ for an exemplary set of conditions. This composition could be explained by assuming that Pluto's building blocks were richer in organic matter than those that formed Titan. It may be assumed that Titan accreted rocks that were compositionally similar to CI chondrites (Tobie et al., 2012; Glein, 2015), while Pluto could have accreted rocks that resembled those in comets Halley and 67P (Kissel and Krueger, 1987; Bardyn et al., 2017). However, present uncertainties are large enough to potentially preclude $NH_3$-$N_{org}$ mixtures from providing a high enough $^{14}N/^{15}N$ ratio. A contribution by primordial $N_2$, which would be a much different scenario than at Titan, would be needed at Pluto if accreted organic matter had a low $^{14}N/^{15}N$ ratio (Figure 3a), or if atmospheric escape significantly decreased the $^{14}N/^{15}N$ ratio in $N_2$ (Figures 2b and 3e). The origin of $N_2$ on Pluto could be related to that on Titan but differ in details, or their origins might be totally different.

How would the above interpretations change if the less conservative lower limit on $^{14}N/^{15}N$ of ~249 from Section 2 were adopted? The major effect would be a reduction in the size of the allowed parameter space (see Figure 3), which may require primordial $N_2$ to be included. However, we cannot say that primordial $N_2$ must be included because $N_{org}$ could provide a $^{14}N/^{15}N$ ratio of up to ~280 (Figures 1 and 3c). I also find that somewhat less $NH_3$ can be accommodated. The upper limit on the fraction of Pluto's $N_2$ that was derived from $NH_3$ would be ~0.34 (Figure 3a, 3b, 3c; cf. Table 2). This reinforces my chief finding that $N_2$ accretion, metamorphism of organic nitrogen, or both processes likely contributed to the origin of Pluto's $N_2$. Finally, the source(s) of observed $N_2$ on Pluto would be made even more different from those on Titan. Isotope ratios along the $N_{org}$-$NH_3$ join would be mostly incompatible with $\left(^{14}N/^{15}N\right)_{N_2,ice} > 249$, except if Pluto accreted organic matter with a light N isotopic composition, and if atmospheric evolution has not appreciably decreased the $^{14}N/^{15}N$ ratio in $N_2$ (e.g., Figure 3c).

It is important to be aware that the above framework for interpreting data on nitrogen isotopes has general utility. The framework itself is not dependent on the particular data that were adopted in this study (i.e., from Lellouch et al., 2017; and Krasnopolsky, 2020b). As work continues, it can be expected that there will be improvements to our understanding of N isotope ratios at Pluto, both from observational and theoretical points of view. If revisions to these data are made, then the present framework can be used to obtain updated results that will provide more detailed and robust insights into how Pluto acquired its molecular nitrogen. There will be lasting value even when data become outdated. This framework can also be useful in aiding interpretations of future measurements of $N_2$ and/or $NH_3$ in potential plumes at Europa by the MAss Spectrometer for Planetary EXploration (MASPEX) instrument onboard Europa Clipper (Waite et al., 2019), and other future measurements of N-bearing compounds in the atmosphere of an ice giant, such as Uranus (Molter et al., 2021), as well as helping to constrain the origin of Triton's $N_2$-rich atmosphere and surface ices (e.g., Rymer et al., 2021).

**4. An integrated perspective on the origin of Pluto's $N_2$ and a look ahead**

This work provides new clarity on the origin of Pluto's $N_2$. While we are not yet able to uniquely identify its source material(s), I have attempted to take a step forward by gaining a deeper understanding of many of the possibilities and their requirements. Table 3 is a compilation of what we have learned through this study. I also found that some contribution (at least ~45% combined) by a primordial $N_2$, organic N, or hybrid source of $N_2$ is implied by a relatively high lower limit on $^{14}N/^{15}N$ in $N_2$-rich ices on Pluto (>197; see Section 2). Ammonia alone is unlikely to be sufficient as a $N_2$ source (Figure 1). The implication is that Pluto had a different cosmochemical origin and geochemical evolution from Titan, where $NH_3$ appears to have played a larger role (Figure 2a). This implication is consistent with the view that impact and photochemical conversion of $NH_3$ to $N_2$ would have been relatively inefficient on Pluto (low impact speeds, cold atmosphere) compared with what is considered plausible on Titan (Atreya et al., 2009; McKinnon et al., 2021). However, these conclusions may be changed if $^{14}$N-$^{15}$N isotopic fractionation on Pluto turns out to be different (Mandt et al., 2017) from the most recent interpretation (Krasnopolsky, 2020b). It is hoped that the present findings can provide focus for future investigations to pursue more diagnostic tests of the origin of N$_2$ on Pluto.

**Table 3.** A synthesis of requirements for Pluto to acquire $N_2$ from single or binary source materials (normal text), and for the indicated materials to provide a low enough $CO/N_2$ ratio (italicized text) and a high enough $^{14}N/^{15}N$ ratio (bold text) to be consistent with present observations (see Glein and Waite, 2018; McKinnon et al., 2021; Sections 1, 3.1, and 3.2; Figures 1 and 3). A three-component source of Pluto's $N_2$ is possible, but is not shown here because the limiting conditions are already captured by binary mixtures.

| Candidate source material(s) of Pluto's $N_2$ | Requirements/notes |
|---|---|
| Primordial $N_2$ | - Low formation temperature (e.g., <30 K) of accreted ices.<br>- *Destruction of CO by subsurface ocean chemistry, or preferential burial of CO relative to $N_2$ in Sputnik Planitia (such burial is deemed unlikely).*<br>- **A primordial $N_2$ source is always consistent with the current constraint on $^{14}N/^{15}N$.** |
| $NH_3$ | - $NH_3$ exposure to a high-energy source at the surface or in the interior.<br>- **Preferential burial of $^{14}N^{15}N$ relative to $^{14}N_2$ in Sputnik Planitia (such burial is deemed unlikely).** |
| Organic N ($N_{org}$) | - Existence of a hot rocky core (≳350°C) and transport of organogenic $N_2$ from the core to the surface.<br>- **If minimal escape-driven isotopic fractionation occurred, then Pluto accreted $N_{org}$ with $^{14}N/^{15}N > 197$.**<br>- **If substantial escape-driven isotopic fractionation occurred, then Pluto accreted $N_{org}$ with $^{14}N/^{15}N > 266$ [a].**<br>- **Alternatively, we cannot impose a requirement on the $^{14}N/^{15}N$ ratio of accreted $N_{org}$ if there has been preferential burial of $^{14}N^{15}N$ relative to $^{14}N_2$ in Sputnik Planitia (such burial is deemed unlikely).** |
| Primordial $N_2$ + $NH_3$ | - Low formation temperature (e.g., <30 K) of accreted ices.<br>- $NH_3$ exposure to a high-energy source at the surface or in the interior.<br>- *Destruction of CO by subsurface ocean chemistry, or preferential burial of CO relative to $N_2$ in Sputnik Planitia (such burial is deemed unlikely).*<br>- **A sufficiently large contribution (>45%) of primordial $N_2$ to Pluto's observed $N_2$ inventory, unless there has been preferential burial of $^{14}N^{15}N$ relative to $^{14}N_2$ in Sputnik Planitia (such burial is deemed unlikely).** |
| Primordial $N_2$ + $N_{org}$ | - Low formation temperature (e.g., <30 K) of accreted ices.<br>- Existence of a hot rocky core (≳350°C) and transport of organogenic $N_2$ from the core to the surface.<br>- *Destruction of CO by subsurface ocean chemistry, or preferential burial of CO relative to $N_2$ in Sputnik Planitia (such burial is deemed unlikely).*<br>- **At least some input of primordial $N_2$ is needed if Pluto accreted $N_{org}$ with $^{14}N/^{15}N < 197$ (for the case of minimal escape-driven isotopic fractionation), or if Pluto accreted $N_{org}$ with $^{14}N/^{15}N < 266$ [a] (for the case of substantial escape-driven isotopic fractionation).**<br>- **Alternatively, we cannot require a primordial $N_2$ input if there has been preferential burial of $^{14}N^{15}N$ relative to $^{14}N_2$ in Sputnik Planitia (such burial is deemed unlikely).** |
| $NH_3$ + $N_{org}$ | - $NH_3$ exposure to a high-energy source at the surface or in the interior.<br>- Existence of a hot rocky core (≳350°C) and transport of organogenic $N_2$ from the core to the surface.<br>- **A $NH_3$ input is permitted if Pluto accreted $N_{org}$ with $^{14}N/^{15}N > 197$ (for the case of minimal escape-driven isotopic fractionation), or if Pluto accreted $N_{org}$ with $^{14}N/^{15}N > 266$ [a] (for the case of substantial escape-driven isotopic fractionation).**<br>- **We cannot restrict the contribution from $NH_3$ if there has been preferential burial of $^{14}N^{15}N$ relative to $^{14}N_2$ in Sputnik Planitia (such burial is deemed unlikely).** |

a Dividing an initial $^{14}N/^{15}N$ ratio of 266 from accreted organic matter by a most likely isotopic enrichment factor for enhanced Jeans escape (1.35; Mandt et al., 2016) yields a present-day $^{14}N/^{15}N$ ratio in $N_2$ ice of 197 (see Section 2).

What should be done next to continue making progress? In the near-term, additional modeling, laboratory work, *New Horizons* data analysis, and comet exploration will all be beneficial. We need to make sure that we understand the magnitude of isotopic fractionation between HCN and $N_2$ in Pluto's atmosphere. This knowledge serves as the foundation of the present paper. Laboratory measurements of $^{14}N$-$^{15}N$ fractionation associated with HCN condensation and aerosol trapping under Pluto atmosphere conditions are needed to test whether the effects of these processes are as large as suggested by Mandt et al. (2017). Also, it needs to be verified that the gas-solid isotope fractionation factor for $N_2$ is not unexpectedly large at Pluto surface temperatures. This will require more experiments. After these experiments have been completed, coupled geochemical-geophysical modeling can be performed to investigate whether vertical variations in the $^{14}N/^{15}N$ ratio of $N_2$ could be supported in Sputnik Planitia. The major conclusions of this paper (see above) would need to be revisited if $^{14}N_2$ were concentrated at the top of the ice sheet. In addition, the history of atmospheric escape of $N_2$ from Pluto should be explored further, building upon the study by Mandt et al. (2016). Future modeling on this topic could consider more complex scenarios with time-varying rates of escape and endogenic input from different $N_2$ sources and their mixtures. It also seems like it may be fruitful to search for spectral signatures of nitriles (organic compounds containing the C≡N group) in reddish regions on Charon (Grundy et al., 2016b). If nitriles are present, then they are likely to have been formed by photolysis of $CH_4$-$N_2$ mixtures (Wong et al., 2015), which may suggest that both $N_2$ and $CH_4$ had significant rates of past escape from Pluto and transport to Charon (Tucker et al., 2015). Lastly, measurements of N isotope ratios in abundant forms of nitrogen (e.g., $N_2$, $NH_3$, $NH_4$-salts, N in organic matter) are needed from multiple comets, particularly Jupiter-family comets (e.g., Wampfler et al., 2022). Such data could be used to improve estimates of initial isotopic ratios on Pluto.

In the more distant future, we will need to send a new mission to Pluto (e.g., Howett et al., 2021). *In situ* measurements of $^{14}N/^{15}N$ in $N_2$ and the $^{36}Ar/N_2$ ratio would provide powerful constraints on the origin of $N_2$ (Glein and Waite, 2018; Miller et al., 2019). *New Horizons* was unable to perform a sensitive search for argon (Mousis et al., 2013; Steffl et al., 2020). The ratio of $^{38}Ar/^{36}Ar$ would independently constrain the influence of escape on atmospheric evolution (e.g., Glein, 2017). Together, these three ratios may enable a unique determination of the source of Pluto's $N_2$. To discern how the origin of $N_2$ relates to that of other volatiles, complementary geochemical data need to be obtained, especially D/H and $^{13}C/^{12}C$ ratios in methane (which may be measurable using the James Webb Space Telescope; Grundy et al., 2011). Additional information that can provide valuable geophysical context includes the outgassed abundance of radiogenic $^{40}Ar$, Pluto's moment of inertia factor, and if Pluto is differentiated, the mean density of its rocky core. Radio tracking of a Pluto orbiter that is equipped with a high-resolution mass spectrometer can deliver these data. If a mission to the Neptune system is launched first, which is very likely (NASEM, 2022), then these same measurements should be prioritized at Triton. This would be enabling to understand how similarities and differences in the origin or evolution of large Kuiper belt objects (past or present) establish their present volatile inventories.

**Acknowledgements**

This study was supported by SwRI internal funding and by the NASA Astrobiology Institute through its JPL-led team entitled *Habitability of Hydrocarbon Worlds: Titan and Beyond*. It was inspired by the many illuminating studies by Toby Owen on the origins of volatiles in the solar system. I thank